\documentclass[12pt]{article}
\usepackage{amsfonts}
\usepackage{epsfig}

\makeatletter

\def\@authoraddress{}
\def\@title{}
\def\title#1{\gdef\@title{{\par\vskip-10pt\Large\bf
\baselineskip20pt\centering\ignorespaces\uppercase{#1}\vskip6pt}}%
\setcounter{table}{0}      \setcounter{figure}{0}
\setcounter{equation}{0}   \setcounter{section}{0}
\setcounter{subsection}{0} \setcounter{subsubsection}{0}
\setcounter{paragraph}{0}
}

\def\authors#1{\expandafter\def\expandafter\@authoraddress\expandafter
{\@authoraddress %
{\dimen0=-\prevdepth \advance\dimen0 by1.5\baselineskip
\nointerlineskip \centering
\vrule height\dimen0 width0pt\relax\ignorespaces\large\sc#1\par
}%
}%
}

\def\addresses#1{\expandafter\def\expandafter\@authoraddress\expandafter
{\@authoraddress{\nointerlineskip\vskip1pc
                 \footnotesize\it\centering\ignorespaces#1\par}}}

\def\nextaddress{\\[2.3pt]}

\def\@maketitle{%
\@title
\ifdim\prevdepth=-1000pt \prevdepth0pt\fi
\@authoraddress
}

\def\maketitle{\par
\begingroup
\let\cite\@bylinecite
\global\@topnum\z@ %
\@maketitle
\endgroup
\def\@thanks{}\def\@authoraddress{}\def\@title{}
}

\def\abstract{\par
\bgroup
\ifdim\prevdepth=-1000pt \prevdepth0pt\fi
\hsize\columnwidth
\leftskip=2em \rightskip\leftskip
\dimen0=-\prevdepth \advance\dimen0 by2pc \nointerlineskip
\noindent\vskip1.5\baselineskip\nointerlineskip\noindent\footnotesize\relax}

\newif\if@firststuff

\def\endabstract{\par
\nointerlineskip \vskip0pt
\noindent \par
\egroup
\hrule depth0pt width0pt
\global\everypar{\global\@firststufffalse}\global\@firststufftrue
}

\renewcommand\section{\@startsection {section}{1}{\z@}%
                                   {-3.5ex \@plus -1ex \@minus -.2ex}%
                                   {2.3ex \@plus.2ex}%
                                   {\normalfont\large\bfseries}}
\renewcommand\subsection{\@startsection{subsection}{2}{\z@}%
                                     {-3.25ex\@plus -1ex \@minus -.2ex}%
                                     {1.5ex \@plus .2ex}%
                                     {\normalfont\large\bfseries}}

\def\1ad{\mbox{\normalsize $^1$}}
\def\2ad{\mbox{\normalsize $^2$}}
\def\3ad{\mbox{\normalsize $^3$}}
\def\4ad{\mbox{\normalsize $^4$}}
\def\5ad{\mbox{\normalsize $^5$}}
\def\6ad{\mbox{\normalsize $^6$}}
\def\7ad{\mbox{\normalsize $^7$}}
\def\8ad{\mbox{\normalsize $^8$}}
\def\adref#1{\mbox{\normalsize $^{#1}$}}
\makeatother

\newcommand{\be}{\begin{equation}}
\newcommand{\ee}{\end{equation}}
\newcommand{\bee}{\begin{eqnarray}}
\newcommand{\eee}{\end{eqnarray}}

\newcommand{\ga}{\alpha}
\newcommand{\gb}{\beta}

\newcommand{\gd}{\delta}

\newcommand{\gep}{\epsilon}

\newcommand{\gs}{\sigma}
\newcommand{\go}{\omega}

\newcommand{\nn}{\nonumber}

\newcommand{\ptl}{\partial}

\renewcommand{\L}{{\cal L}}
\renewcommand{\P}{{\cal P}}
\newcommand{\K}{{\cal K}}
\newcommand{\D}{{\cal D}}
\newcommand{\R}{{\cal R}}

\renewcommand{\S}{{\cal S}}

\newcommand{\vac}
{|0\rangle}

\newcommand{\vphi}
{|\Phi(a^+,x)\rangle}

\begin{document}
\raggedbottom

\title{$3d$ Conformal Higher Spin Symmetry in
$2+1$ Dimensional Matter Systems }

\authors{O.V.~Shaynkman,\adref{1}}

\addresses{I.E.Tamm Department of Theoretical Physics, Lebedev Physical
Institute,\\
Leninsky prospect 53, 119991, Moscow, Russia \nextaddress\1ad
shayn@lpi.ru}

\maketitle

\begin{abstract}
The symmetry algebra of massless fields living on the
3-dimensional conformal boundary of $AdS_4$ is shown to be
isomorphic to $3d$ conformal higher spin algebra ($AdS_4$ higher
spin algebra). A simple realization of this algebra on the free
flat $3d$ massless matter fields is given in terms of an auxiliary
Fock module.
\end{abstract}

\section{Introduction}
The first example of the full interacting higher spin theory was
given in 1990 by M. Vasiliev  \cite{1} (see also \cite{2,3,3_} for
reviews). It is formulated in $AdS_4$ space and describes an
interaction of the fields of all spins, which is invariant under
the $AdS_4$ higher-spin gauge symmetry algebra \cite{6,8}. It was
then conjectured by Fradkin and Linetsky \cite{FL} that 3d
conformal higher spin algebras have to be identified with the
$AdS_4$ higher spin algebra. One motivation of this work is to
check whether the symmetries of the massless fields living on the
$3$-dimensional conformal boundary of $AdS_4$ do form the  3d
conformal higher spin algebra. It will be shown that this is
indeed the case.

Another motivation comes from Maldacena conjecture on AdS/CFT
correspondence \cite{5} that suggests a duality between the theory
of gravity in the $AdS_4$ bulk space and 3-dimensional conformal
theory on the boundary of $AdS_4$. Since $AdS_4$ higher spin
theory contains supergravity it seems natural to speculate that
$AdS_4$ higher spin theory should have some dual 3-dimensional
theory possessing the $AdS_4$ higher spin symmetry algebra as 3d
conformal higher spin algebra. It is therefore interesting to
realize the 3d conformal higher spin algebra on the 3-dimensional
matter fields (scalar and spinor), which are the only conformal
fields in 3-dimensional higher spin theory. This realization is
the main topic of this talk. It is obtained using the method of
unfolded formulation, which is very suitable  for the analysis of
symmetries of  dynamical systems. In this presentation we
summaries the main ideas and results. More details can be found in
the original work \cite{7}, where not only flat but also $AdS_3$
matter system is considered and the duality between non-unitary
field-theoretical representations of the conformal algebra and the
unitary (singleton--type) representations \cite{4} of the $3d$
conformal algebra $sp(4,\R)$ is found in the form of a certain
Bogolyubov transform.

Let us note that recently in the interesting paper \cite{KP}
Klebanov and Polyakov suggested that the full nonlinear $AdS_4$
higher spin theory is dual to the large $N$ limit of the conformal
$3d$ vector model which is a nonlinear deformation of the system
considered here. Hopefully, an appropriate deformation of the
system within our scheme may help to check this conjecture.

\section{$AdS_4$ higher spin algebra and its oscillator realization}
The infinite dimensional superextension $f$ of the ordinary
$AdS_d$ algebra $so(d-1,2)$ is called $AdS_d$ higher spin algebra
if \cite{6}:\\
1. $f$ is graded, i.e. $f$ decomposes into the infinite direct sum
\be\label{graded_algebra} f=\oplus_{l=-\infty}^{\infty}f_l \ee of
finite-dimensional subspaces $f_l$, which are called levels, and
$[f_l,f_q]\subseteq f_{l+q}$\\
2. every level $f_l$ forms a finite-dimensional representation of
the Lorentz subalgebra $so(d-1,1)\subset so(d-1,2)$\\
3. and, finally, $f$ obey the admissibility condition \cite{11} of
possessing massless unitary representation with the same spectra
of spins as predicted by the structure of gauge fields originating
from $f$.\\
The simplest version of the $AdS_4$ higher spin algebra ($3d$
conformal higher spin algebra) can be constructed with the help of
the oscillator realization of the $AdS_4$ algebra \cite{8}.

Consider oscillators $a_\ga$ and $a^+_\gb$ ($\ga,\gb=1,2$ are
spinorial indices) with the commutation relations
\be
[a_\ga,a^+_\gb ]=\gep_{\ga\gb}\,,\qquad [a_\ga,a_\gb ]=[ a^+_\ga,
a^+_\gb]=0\,.
\label{oscillators}
\ee
Here $\gep_{\ga\gb}$ is antisymmetric charge conjugation matrix
$\gep_{12}=\gep^{12}=1$ that rises and lowers spinorial indices
\be
a^\ga=a_\gb\gep^{\ga\gb}\,,\qquad a_\ga=a^\gb\gep_{\gb\ga}\,.
\ee
The bilinear combinations of oscillators (\ref{oscillators}) form
the algebra $sp(4,\R)\sim so(3,2)$. The conformal basis of
$so(3,2)$ is given by the formulas
\bee\label{so_basis}
&\L_{nm}=-\frac{1}{4}\gep_{nmk}\gs^{k\ga\gb}(a_\ga a^+_\gb+
a^+_\gb a_\ga)\,,\nn\\
&\P_n=\frac{1}{2}\gs_n{}^{\ga\gb}a_\ga a_\gb\,,
\qquad \K_n=\frac{1}{2}\gs_n{}^{\ga\gb}a^+_\ga a^+_\gb\,,\\
&\D=\frac{1}{4}(a_\ga a^{+\ga}+a^{+\ga}a_\ga)\nn\,,
\eee
where $\L_{nm}$ are generators of Lorentz subalgebra
$so(2,1)\subset so(3,2)$ while $\P_n\,,$ $\K_n$ and $\D$ are
generators of momenta, special conformal transformations and
dilatation. Here $n,m,k=1,2,3$ and $\gs_n^{\ga\gb}=\gs_n^{\gb\ga}$
are the unit matrix and the first and the third Pauli matrixes,
respectively.

Considering  also linear combinations of
oscillators (\ref{oscillators}) we extend the algebra
$so(3,2)$ to the superalgebra $osp(1|4,\R)$ with supergenerators
\be
\S_\ga=a^+_\ga\,,\qquad Q_\gb=a_\gb\,.
\ee

The $AdS_4$ higher spin algebra \cite{6,8} is defined as a further
generalization of this construction. Let $A$ be the associative
algebra, enveloping the oscillator commutation relations
(\ref{oscillators}), i.e. $A$ contains all possible combinations
of oscillators. Let Lie($A$) be Lie superalgebra built from $A$
with use of the supercommutator \be [f,g]_\pm =fg-(-1)^{\pi (f)\pi
(g)}gf\,, \ee where $f(a,a^+)$, $g(a,a^+)\in A$ are some monomials
from $A$ and $\pi(f)$ is the parity function which counts the
oddness of the oscillators in $f(a,a^+)$ \be
f(a,a^+)=(-1)^{\pi(f)}f(-a,-a^+)\,. \ee Following the general
notation of  \cite{8_} let $hgl(1;1|4)$ denote Lie algebra
Lie($A$). The $AdS_4$ higher spin algebra ($3d$ conformal higher
spin algebra), denoted by $hu(1;1|4)$, is a real part of
$hgl(1;1|4)$ which is singled out by the condition \be
-(i)^{\pi(f)}f(ia,ia^+)=f(a,a^+) \ee for $f\in hgl(1;1|4)$.

It can be easily seen that the algebra $hu(1;1|4)$ fits the first
and the second parts of the definition of the $AdS_4$ higher spin
algebra given at the beginning of the section. Really it is graded
(\ref{graded_algebra}), with the levels $hu(1;1|4)_l$ consisting
from those elements of $hu(1;1|4)$ that \be \# \mbox{ of }a^+ -\#
\mbox{ of }a=l \ee and every level is closed under the adjoint
action of  $\L_{nm}$. In \cite{11} it was sown that $hu(1;1|4)$
also satisfy admissibility condition (the third part of the
definition).

\section{Unfolded formulation of dynamical systems:\\
Massless Scalar Example}
The general properties of the unfolded
formulation of dynamical systems were studied in \cite{9,10}. The
key observation is that any free dynamical system can be
reformulated in a form of zero curvature and covariant constancy
conditions
\bee
&\label{zero_curvature_eq} d \go(x) =\go(x)\wedge\go(x)\,,\\
&\label{covariant_constancy_eq} d \vphi =t(\go(x))\vphi\,, \eee
where $\go(x)$ is connection 1-form taking values in the gauge
symmetry algebra of the considered dynamical system and $\vphi$ is
some vector from the representation $t(\go(x))$ of this algebra.
For such a formulation to work the vector $\vphi$ should contain
usually infinite set of auxiliary fields.

In what follows we require the connection 1-form $\go(x)$
to take values in $AdS_4$ higher spin algebra $hu(1;1|4)$ and
$t(\go(x))$ to be a Fock representation of $hu(1;1|4)$, generated
by $a^+_\ga$ from the Fock vacuum $\vac$ that is annihilated by
$a_\ga$
\be\label{Fock_vacuum}
a_\ga\vac=0\,.
\ee

Let us illustrate the properties of the unfolded formulation
(\ref{zero_curvature_eq}), (\ref{covariant_constancy_eq}) of the
dynamical systems by the example of the free massless flat
3-dimensional Klein-Gordon equation $\Box c(x)=0$. To this end we
introduce infinitely many auxiliary fields, which are  totally
symmetric multispinors of all even ranks
\be
c_{\ga_1\ga_2}(x)\,,c_{\ga_1\ga_2\ga_3\ga_4}(x)\,,\ldots\,.
\ee
Consider the infinite system of differential equations  for these
fields and the original scalar field $c(x)$
\bee
&\label{unfolded_K_G_1}\ptl_n c(x)=\frac{1}{2}\gs_n{}^{\ga_1\ga_2}
c_{\ga_1\ga_2}(x)\,,\\
&\label{unfolded_K_G_2}\ptl_n
c_{\ga_1\ga_2}(x)=\frac{1}{2}\gs_n{}^{\ga_3\ga_4}
c_{\ga_1\ga_2\ga_3\ga_4}(x)\,,\\
&\cdots\nn\\
&\label{unfolded_K_G_2k}\ptl_n
c_{\ga_1\ldots\ga_{2k}}(x)=\frac{1}{2}
\gs_n{}^{\ga_{2k+1}\ga_{2k+2}}c_{\ga_1\ldots\ga_{2k+2}}(x)\,,\\
&\cdots\nn
\eee
Since the matrixes $\gs_n{}^{\ga\gb}$ are reversible
\be
\gs_n{}^{\ga\gb}\gs_m{}_{\ga\gb}=2\eta_{nm}
\ee
we can express all the higher multispinors
$c_{\ga_1\ldots\ga_{2k}}(x)$ ($k\geq 1$) via the derivatives of
the dynamical field $c(x)$ as
\bee
&\label{constrain_1}c_{\ga_1\ga_2}(x)=\gs^n{}_{\ga_1\ga_2}\ptl_n
c(x)\,,\\
&\label{constrain_2}c_{\ga_1\ga_2\ga_3\ga_4}(x)
=\gs^{n_1}{}_{\ga_1\ga_2}
\gs^{n_2}{}_{\ga_3\ga_4}\ptl_{n_1} \ptl_{n_2}c(x)\,,\\
&\cdots\nn\\
&\label{constrain_k}c_{\ga_1\ldots\ga_{2k}}(x)
=\gs^{n_1}{}_{\ga_1\ga_2}\cdots
\gs^{n_k}{}_{\ga_{2k-1}\ga_{2k}}\ptl_{n_1}\cdots \ptl_{n_k}c(x)\,,\\
&\cdots\nn
\eee
Note that the left hand sides of these expressions are required to
be totally symmetric. This has a consequence that the field $c(x)$
has to satisfy the Klein-Gordon equation
\be\label{K_G}
\Box c(x)=0
\ee
being the consistency condition for this system. Therefore,
the system (\ref{unfolded_K_G_1}), (\ref{unfolded_K_G_2}),
(\ref{unfolded_K_G_2k}) is equivalent to the infinite set of
constraints (\ref{constrain_1}), (\ref{constrain_2}),
(\ref{constrain_k}) for all higher multispinors and the
Klein-Gordon equation for the dynamical field $c(x)$.

The analogous system of equations for the spinor field
$c_\ga(x)$ involves all odd rank auxiliary multispinors
$c_{\ga_1\ldots\ga_{2k+1}}(x)$ ($k\geq 1$). It encodes the free
massless Dirac equation
\be\label{Dirac}
\ptl_n\gs^n{}_{\ga}{}^{\gb}c_\gb(x)=0\,.
\ee
Thus we  conclude that Klein-Gordon and Dirac equations
(\ref{K_G}), (\ref{Dirac}) are concisely encoded as consistency
conditions of the system
\be
\label{unfolded_K_G_and_Dirac}\ptl_n c_{\ga_1\ldots\ga_{l}}(x)
=\frac{1}{2}\gs_n{}^{\ga_{l+1}\ga_{l+2}}c_{\ga_1\ldots\ga_{l+2}}(x)
\qquad l=0,\ldots,\infty\,,
\ee
where $c(x)$ and $c_\ga(x)$ are dynamical fields and totally
symmetric multispinors $c_{\ga_1\ldots\ga_l}(x)$ ($l\geq 2$) are
axillary fields expressed via higher derivatives of the dynamical
fields.

In fact, the system (\ref{unfolded_K_G_and_Dirac}) is a special
case of the general system (\ref{zero_curvature_eq}),
(\ref{covariant_constancy_eq}) when $\go(x)$ is a flat connection
of particular form
\be
\go(x)=\go_0(x)=\frac{1}{2}dx^n\gs_n{}^{\ga\gb}a_\ga a_\gb
\ee
and
\be
\vphi=\sum_{l=0}^\infty \frac{1}{l!}c_{\ga_1\ldots\ga_l}(x)a^{+\ga_1}
a^{+\ga_l}\vac\
\ee
is some element of an axillary Fock space (\ref{Fock_vacuum})
playing the role of a generating function for the $c(x)$,
$c_\ga(x)$ and all axillary fields $c_{\ga_1\ldots\ga_l}(x)$
($l\geq 2$).

It is important that any  system that can be rewritten in the form
(\ref{zero_curvature_eq}), (\ref{covariant_constancy_eq}) (in
particular, the system (\ref{unfolded_K_G_and_Dirac})) is
invariant under the following transformations
\bee\label{gauge_transformations}
&\gd\go(x)=d\gep(x)-[\go(x),\gep(x)]\,,\nn\\
&\gd\vphi=\gep(x)\vphi
\eee
with the gauge symmetry parameter $\gep(x)$. The different choices
of the connection $\go(x)$ in (\ref{zero_curvature_eq}),
(\ref{covariant_constancy_eq}) correspond to the different choices
of space-time coordinates in the differential equations under
consideration (like Klein-Gordon and Dirac equation). Once
some solution of the zero curvature equation
(\ref{zero_curvature_eq}) is fixed, the gauge symmetry
(\ref{gauge_transformations}) breaks down to the global symmetry
\bee
&\label{connection_fixed}\gd\go(x)=0\,,\\
&\label{global_symmetry}\gd\vphi=\gep_0(x)\vphi \eee with the
symmetry parameter $\gep_0(x)$ keeping the connection $\go(x)$
invariant and thus satisfying the differential equation
\be\label{eq_for_global_symmetry_parameter}
 d\gep_0(x)-[\go_0(x),\gep_0(x)]=0\,.
\ee
We thus obtain that the system (\ref{unfolded_K_G_and_Dirac}) and,
most important, Klein-Gordon and Dirac equations encoded by it are
invariant under the transformations (\ref{global_symmetry}). So
the unfolded formulation allows one to write down the dynamical
equations in a manifestly symmetric form.

Also such an approach allows one to write generic solution of
the system (\ref{zero_curvature_eq}), (\ref{covariant_constancy_eq})
as well as the solution of the equation
(\ref{eq_for_global_symmetry_parameter}) for the global symmetry
parameter $\gep_0(x)$. These solutions are
\bee
&\label{solution_of_zero_curvature_eq}\go_0(x)=-g(x)^{-1}d g(x)\,,\\
&\label{solution_of_covariant_constancy_eq}
\vphi=g(x)^{-1}|\Phi_0\rangle\,,\\
&\label{solution_of_eq_for_global_symmetry_parametr}
\gep_0(x)=g(x)^{-1}\xi g(x)\,. \eee Here $g(x)\in A$ is some
invertible element constructed from oscillators, which give rise
to $\go_0(x)$ in a pure gauge form
(\ref{solution_of_zero_curvature_eq}) and $\xi$ is an arbitrary
element of the $AdS_4$ higher spin algebra ($3d$ conformal higher
spin algebra) $hu(1;1|4)$, playing a role of the initial data for
the differential equation (\ref{eq_for_global_symmetry_parameter})
$\gep_0(x_0)=\xi$ for $x_0:$ $g(x_0)=1$. The formula
(\ref{solution_of_eq_for_global_symmetry_parametr}) realizes the
isomorphism between algebra of the symmetries of the free flat
massless 3-dimensional Klein-Gordon and Dirac equations and 3d
conformal higher spin algebra $hu(1;1|4)$. Moreover, this formula
allows us to obtain the particular realization of $hu(1;1|4)$ on
3-dimensional massless scalar and spinor.

\section{Realization of $hu(1;1|4)$ on 3d matter system}
Let us come back to the  system of interest
(\ref{unfolded_K_G_and_Dirac}). From the equation
(\ref{solution_of_zero_curvature_eq}) it can be easily seen that
$g(x)\in A$, giving rise to flat connection
$\go_0(x)=\frac{1}{2}dx^n\gs_n{}^{\ga\gb}a_\ga a_\gb$, can be
taken in the form \be g_0(x)=\exp
(-\frac{1}{2}x^n\gs_n{}^{\ga\gb}a_\ga a_\gb). \ee Using
(\ref{solution_of_eq_for_global_symmetry_parametr}) and
(\ref{global_symmetry}) we can represent $hu(1;1|4)$ on the set of
all (dynamical and axillary) fields $c_{\ga_1\ldots\ga_l}(x)$
$l\geq 0$. Then taking into account the constrains
(\ref{constrain_1}), (\ref{constrain_2}), (\ref{constrain_k}) we
can finally realize $hu(1;1|4)$ as differential operators acting
on the matter fields $c(x)$ and $c_\ga(x)$.  For the simple
low-order polynomials $\xi$ this can be done "by hand" using the
Campbell-Hausdorf formula \be
e^Ae^B=e^{A+B+\frac{1}{2}[A,B]+\cdots}\,, \ee with the only first
three nonzero terms. In particular, for the conformal subalgebra
$so(3,2)\subset hu(1;1|4)$ we arrive at well known result \bee
&&\!\!\!\!\!\!\!\!\!\L_{nm}c(x)=(x_m\ptl_n-x_n\ptl_m)c(x)\,,\nn\\
&&\!\!\!\!\!\!\!\!\!
\L_{nm}c_\ga(x)=\Big((x_m\ptl_n-x_n\ptl_m)\gd_\ga^\gb+
\frac{1}{2}\gep_{nmk}\gs^{k\gb}{}_\ga\Big) c_\gb(x)\,,\nn\\
&&\!\!\!\!\!\!\!\!\!\P_n c(x)=\ptl_n
c(x)\,,\qquad\qquad\qquad\qquad\;
\P_n c_\ga(x)=\ptl_n c_\ga(x)\,,\nn\\
&&\!\!\!\!\!\!\!\!\!\K_n c(x)=(x_n+2x_n x^k\ptl_k-x^2\ptl_n )c(x)\,,\\
&&\!\!\!\!\!\!\!\!\!\K_n c_\ga(x)= \Big((2x_n+2x_n x^k
\ptl_k-x^2\ptl_n)\gd_\ga^\gb-
\gep_{nmk}x^m\gs^{k\gb}{}_\ga\Big) c_\gb(x)\,,\nn\\
&&\!\!\!\!\!\!\!\!\!\D c(x)=(\frac{1}{2}+x^k \ptl_k) c(x)\,,\qquad
\qquad\;\;\;\D c_\ga(x)=(1+x^k\ptl_k) c_\ga(x)\nn\,.
\eee

For the general case the computations are less trivial and involve
the technique of symbols of operators. The final result for a
general element of $hu(1;1|4)$ represented with use of the
generating function $\exp(a^{+\ga}h_\ga+a_\ga h^{+\ga})$ is of the
following form \cite{7} \be \!\!\delta\vphi \!=\! \exp\!
\Big(\frac{1}{2}x^{\ga\ga}h_\ga h_\ga+ a^{+\ga}h_\ga
+\frac{1}{2}h^{+\ga}h_\ga \Big) \Big|\Phi \Big(
a^{+\ga}+x^{\ga\gb}h_\gb+h^{+\ga},x\Big) \Big\rangle\,.
\label{fdc} \ee

\section{Acknowledgement}
This research was supported in part by INTAS, Grant No.99-1-590,
the RFBR Grant No.00-15-96566, the RFBR Grant No.02-02-17067 and
the RFBR Grant No.02-02-06512.


\begin{thebibliography}{9}
\bibitem{1} M.A. Vasiliev, {\it Phys.Lett.}  {\bf B 243} (1990) 378,
{\it Phys.Lett.} {\bf B 285} (1992) 225.
\bibitem{2} M.A. Vasiliev,
Contributed article to Golfand's Memorial Volume, ed. by
M.~Shifman, {\tt hep-th/9910096}.
\bibitem{3}
M.A.~Vasiliev, {\it Int.J.Mod.Phys.} {\bf D 5} (1996) 763.
\bibitem{3_}
E. Sezgin, P. Sundell, {\tt hep-th/0211113}.
\bibitem{6}  E.S. Fradkin, M.A. Vasiliev,
{\it Docl.Akad.Nauk.\/} {\bf 29} (1986) 1100, {\it Ann.Phys.\/}
{\bf 177} (1987) 63.
\bibitem{8} M.A. Vasiliev, {\it Fortschr.Phys.} {\bf 36} (1988) 33,
{\it Nucl.Phys.\/} {\bf B 301} (1988) 26.
\bibitem{FL} E.S. Fradkin, V.Ya. Linetsky {\it Ann.Phys.} (N.Y.)
{\bf 198} (1989) 252.
\bibitem{5} J. Maldacena, {\it
Adv.Theor.Math.Phys.} {\bf 2} (1998) 231, {\it Int.J.Theor.Phys.}
{\bf 38} (1998) 1113,
{\tt hep-th/9711200};\\
S.S. Gubser, I.R. Klebanov,  A.M. Polyakov, {\it Phys.Lett.}  {\bf
B 428} (1998) 105,
{\tt hep-th/9802109};\\
E. Witten, {\it Adv.Theor.Math.Phys.} {\bf 2} (1998) 253,
 {\tt hep-th/9802150}.
\bibitem{7} O.V. Shaynkman, M.A. Vasiliev, {\it Teor.Math.Phys.\/}
{\bf 128} (2001) 1155, {\tt hep-th/0103208}.
\bibitem{4} P.A.M.
Dirac, {\it J.Math.Phys.} {\bf 4} (1963) 901.
\bibitem{KP} I.R. Klebanov, A.M. Polyakov, {\tt hep-th/0210114}.
\bibitem{11} S.E. Konstein, M.A. Vasiliev, {\it Nucl.Phys.\/}
{\bf B 312} (1989) 402.
\bibitem{8_} S.E. Konstein, M.A. Vasiliev,
{\it Nucl.Phys.\/} {\bf B 331} (1990) 475.
\bibitem{9}   M.A. Vasiliev, {\it Ann.Phys.} (N.Y.) {\bf 190} (1989) 59.
\bibitem{10}  M.A. Vasiliev, {\it Phys.Rev.} {\bf D66} (2002),
{\tt hep-th/0106149}.
\end{thebibliography}
\end{document}